\documentclass{emulateapj}
\usepackage{apjfonts}

\shorttitle{Reheating White Dwarfs in AM CVn Binaries}
\shortauthors{Bildsten, Townsley, Deloye and Nelemans }
\submitted{Accepted to the Astrophysical Journal}
\newcommand{\timav}{\langle \dot M \rangle}

\begin{document}

\title{The Thermal State of the  Accreting White Dwarf in AM Canum Venaticorum Binaries}

\author{Lars Bildsten}
\affil{Kavli Institute for Theoretical Physics and Department of Physics\\
Kohn Hall, University of California, Santa Barbara, CA 93106;
bildsten@kitp.ucsb.edu}

\author{Dean M. Townsley}
\affil{Department of Astronomy and Astrophysics, 5640 South Ellis Avenue, University of Chicago, Chicago, IL 60637; townsley@uchicago.edu}

\author {Christopher J. Deloye}
\affil{
Department of Physics \& Astronomy, Northwestern University, Evanston, IL 60208; cjdeloye@northwestern.edu}

\author{Gijs Nelemans} 
\affil{Institute of Astronomy, Madingly Road, CB3 0HA, Cambridge, UK, and
Department of Astrophysics, Radboud University
Nijmegen, Toernooiveld 1, NL-6525 ED, The Netherlands;  nelemans@astro.ru.nl}

\begin{abstract}
 
  We calculate the heating and cooling of the accreting white dwarf
(WD) in the ultracompact AM Canum Venaticorum (AM CVn) binaries and
show that the WD can contribute significantly to their optical and
ultraviolet emission.  We estimate the WD's effective temperature,
$T_{\rm eff}$, using the optical continuum for a number of observed
binaries, and show that it agrees well with our theoretical
calculations. Driven by gravitational radiation losses, the time
averaged accretion rate, $\timav$, decreases monotonically with
increasing $P_{\rm orb}$, covering six orders of magnitude.  If the
short period ($P_{\rm orb}<10$ min) systems accrete at a rate
consistent with gravitational radiation via direct impact, we predict
their unpulsed optical/UV light to be that of the $T_{\rm eff}>50,000$
K accreting WD.  At longer $P_{\rm orb}$ we calculate the $T_{\rm
eff}$ and absolute visual magnitude, $M_V$, that the accreting WD will
have during low accretion states, and find that the WD naturally
crosses the pulsational instability strip.  Discovery and study of
pulsations could allow for the measurement of the accumulated helium
mass on the accreting WD, as well as its rotation rate.  Accretion
heats the WD core, but for $P_{\rm orb}>40 $ minutes, the WD's $T_{\rm
eff}$ is set by its cooling as $\timav$ plummets. For the two long
period AM CVn binaries with measured parallaxes, GP Com and CE 315, we
show that the optical broadband colors and intensity are that expected
from a pure helium atmosphere WD.  This confirms that the WD
brightness sets the minimum light in wide AM CVn binaries, allowing
for meaningful constraints on their population density from deep
optical searches, both in the field and in Globular Clusters.

\end{abstract}


\section{Introduction}

  The AM Canum Venaticorum (AM CVn) class of ultracompact binaries
provides an opportunity for probing the accretion-induced heating of a
white dwarf (WD) over six orders of magnitude in the time-averaged
accretion rate, $\timav$. In these very tight ($P_{\rm orb}< 80 $ min)
binaries (see Warner 1995 for an overview), the donor star is a very
low mass ($M<0.1M_\odot$) helium WD and the accretor is either a He or
a C/O WD. Though there are a few distinct AM CVn formation scenarios
(e.g. Nelemans et al. 2001 and references therein), they all evolve as
mass-transferring binaries driven by the loss of angular momentum from
gravitational radiation (e.g. Deloye, Bildsten \& Nelemans 2005).

 This paper is an outgrowth of Townsley and Bildsten's (2004,
hereafter TB) work on the reheating of WDs in H-accreting cataclysmic
variables. However, the case of WD reheating in AM CVn binaries is
different because the accreted material is pure helium, making flashes
an isolated rather than constant occurence, and the binary evolution
proceeds only under angular momentum loss by gravitational radiation,
leading to a plummeting $\timav$ with increasing $P_{\rm orb}$.  We
have found that the WD is heated in the earliest phases of binary
evolution and cools at later times.  There is a critical decoupling
accretion rate of $\dot M_{\rm dec}\approx (1-3)\times 10^{-10}
{M_{\odot} \ \rm yr^{-1}}$ below which the reheated WD will cool even
while accreting.  We find the amount of time that has passed since the
binary had this $\timav$ by using the evolutionary calculations of
Deloye et al. (2005).  Simple cooling after this point predicts the WD
luminosity, as well as the expected WD $T_{\rm eff}$ and absolute
visual magnitude, $M_V$ (assuming, of course, a WD mass, $M$, and
radius, $R$).  We show that the accreting WD can become the dominant
broadband optical light source in those binaries with orbital periods
in excess of $\approx 40$ minutes.

 In \S 2, we discuss the physics of the WD heating due to accretion
and discuss the uncertainty in the total helium layer mass expected on
the WD.  Since the WD luminosity, $L$, is nearly always less than the
accretion luminosity, $L_{\rm acc}=GM\timav/R$, we explain in \S 3 how
the binary's behavior allows the WD to be directly detected at
different orbital periods. At early times, during direct impact
accretion (e.g. Marsh and Steeghs 2002), the unpulsed UV/optical light
is from the hot WD, whereas at longer orbital periods, the disk is
often quiescent, allowing for the underlying WD to dominate the
optical continuum of a number of these binaries. We also speculate on
what could be learned if one of these WDs was caught in the DB
pulsational instability strip. We make direct comparisons to the
observations in \S 4, and conclude in \S 5.

\section{The Physics of White Dwarf Heating in AM CVn Binaries} 

 The ability for accretion to heat the deep interior of a WD has been
discussed extensively (Sion 1991; TB). Indeed, accretion of pure
helium at rapid rates $\dot M>10^{-7}\ M_\odot \ {\rm yr^{-1}}$ is
known to flash unstably and was studied as a Type Ia progenitor
scenario (e.g. Nomoto \& Sugimoto 1977; Nomoto 1982; Woosley \& Weaver
1994).  Though eventually ruled out as a major mechanism for Type Ia
supernovae (Solheim \& Yungelson 2005), AM CVn binaries accrete pure
helium for a long enough time to reheat the accreting WD.  An
additional feature of these binaries is the large dynamic range in the
time averaged accretion rate, $\timav$, onto the WD, allowing us to
test the theories of re-heated WDs, as discussed in \S 3 \& 4.

Here we calculate the accretor's thermal evolution along with that of
the binary for several AM CVn systems that represent the expected
ranges for this population. Motivated by the expected mass
distribution of He WD donors that evolve into AM CVn systems
(Nelemans et al.  2001), we chose an initial donor mass of $0.24 M_\odot$.  
The thermal state of the donor is very important for
setting the relationship between $\timav$ and $P_{\rm orb}$ (Bildsten
2002; Deloye \& Bildsten 2003; Deloye et al. 2005). For this study we
chose two thermal states found by setting the initial entropy of the
donor so that the system's $P_{\rm orb}$ at contact is greater than
the $P_{\rm orb}$ at contact of 30\% and 90\% of the systems in the
Deloye et al. (2005) AM CVn population evolution calculation.  This
provides $\timav$ as a function of time, and we then use the
quasi-static envelope methods of TB to heat the accreting 
WD.  For simplicity, this evolution is performed at constant $M$ and
with a constant helium layer.  We chose two masses $M=0.65M_\odot$
and $1.05M_\odot$, each case with a $0.05M_\odot$ accreted helium
shell (justified below). Initial $M$ values for the binary evolution were chosen such
that $M(t)$ matched that of the thermal calculations at late times
(long $P_{\rm orb}$) to ensure the correct $P_{\rm orb}$-time
relation.

\begin{figure}
\plotone{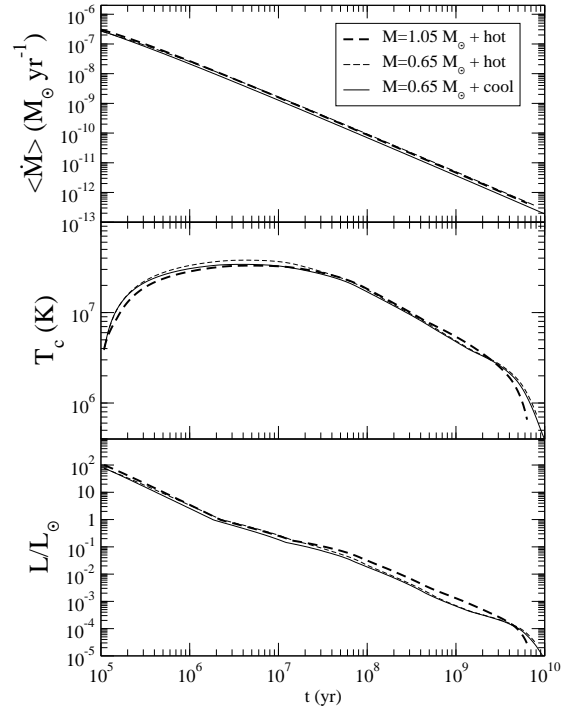}
\figcaption{
\label{fig:theory}
The time averaged mass transfer rate $\timav$, WD core temperature, 
$T_c$, and WD luminosity, $L$,  as a function of time since the onset
of mass transfer for donors with fixed entropy. The thin (thick) lines
show the evolution for a $0.65M_\odot$ ($1.05M_\odot$) WD,
0.05$M_\odot$ of which is a surface helium layer. For the lower mass
we show two curves with low (solid, cool) and high (dashed, hot) donor 
entropy (from Deloye et al. 2005).  There are four epochs:
(1) (not shown) rapid accretion with helium shell flashes (2) $T_c$
rises at early times when $\timav$ becomes low enough that shell
flashes are suppressed and the accreted helium layer can build up, (3)
$\timav$ falls enough that $T_{c,\rm eq}\approx T_c$ and core heating
ceases, (4) $\timav$ becomes low enough that the cooling of the core
dominates the outgoing luminosity.
}
\end{figure}
Using these $\timav$ trajectories (see Figure 1), we identified four
relevant regimes for the state of the accreting WD.  Chronologically
since initiation of mass transfer (and in order of decreasing
$\timav$), these are: (1) rapid accretion and thermally unstable He
ignition ($<{\rm few}\times 10^5 {\rm yr}$), (2) core heating during a
phase in which $L$ is determined from ``compressional'' heating in the
WD envelope ($\sim 3\times 10^5-3\times 10^6 {\rm yr}$), (3) a
transition period where the core temperature is near an equilibrium
value set by $\timav$ ($3\times 10^6-10^8{\rm yr}$), and (4) the WD
cooling regime ($>10^8 \ {\rm yr}$) where $L$ is determined by the
time, $\Delta t$, since the heating stopped at a previous epoch where
$\timav\sim \dot M_{\rm dec}$, the decoupling mass-transfer rate.  We
now describe each regime.

  At the very earliest times, when $\timav\gtrsim 10^{-7}M_\odot\ {\rm
yr^{-1}}$, the accreted helium is compressed rapidly and ignites in a
thermally unstable manner, leading to recurrent weak explosions
(Fujimoto \& Sugimoto 1982).%
\footnote{Extremely high accretion rates for which a red giant
envelope can be supported, $\timav \gtrsim L_{\rm He\
Edd}/\epsilon_{n} = 7\times 10^{-6}M_\odot\ {\rm yr}^{-1}(M/M_\odot)$
(where $\epsilon_{n}=0.6\ {\rm MeV}/$amu is the energy released from
helium fusion) only appear at the brief ($<10^3$ yr) and poorly
understood onset of accretion.}  It is uncertain how much of the
accreted material is ejected (see Kato \& Hachisu 1999).  As $\timav$
drops, the compression happens more slowly so that the envelope
temperature is lower, and the helium shell can become quite
large. Thermal instabilities at this stage likely lead to detonations
(Fujimoto \& Sugimoto 1982), such that the last thermonuclear
instability might be the most violent. However, once $\timav< 2\times
10^{-8}M_\odot \ {\rm yr^{-1}}$, the required unstable He ignition
mass for $0.6-1.1 M_\odot$ WDs (Nomoto 1982; Limongi \& Tornambe 1991;
Goriely et al. 2002) exceeds the donor mass.  Thus once a system
reaches an orbital period of 10 minutes, it is unlikely to undergo
further unstable helium ignition events.  That is, $\timav$ drops
quickly enough that once $\timav$ is low enough for a strong flash,
the accreted layer never becomes thick enough for it to occur. Over
the remaining course of the binary's evolution, about $0.1 M_\odot$ is
accreted by the WD, motivating us to use a fixed accreted layer mass
of $0.05 M_\odot$ to calculate the thermal evolution of the accretor.
Because $L$ depends only logarithmically on the accreted layer mass
(TB), using this fixed value has only a small impact, especially when
compared to the large variations in $L$ from the orders of magnitude
variation in $\timav$.

 The period of time during which $\timav > 10^{-7} M_\odot$ yr$^{-1}$
is so short ($\sim 10^5$ yr) that the accretor's core is thermally
unperturbed from its initial state. However, once the helium layer
begins to accumulate, thermal diffusion both out of the surface
(setting the exiting luminosity, $L$) and into the core (acting to
increase the temperature there, $T_c$) occurs.  We thus begin evolving
the WD's thermal state $10^5$ yr after contact. Although the
compressional energy release depends weakly on the accumulated He
layer mass (i.e.  logarithmically, TB), the $L$-$T_c$ relation which
determines the cooling is essentially independent of it.  The
evolution is shown in Figure 1 and exhibits all of the epochs (2)-(4)
including reheating, decoupling and cooling.  The heat capacity and
crystallized fraction of the core was calculated using the equation of
state of Potekhin \& Chabrier (2000), and the late-time $L$-$T_c$
relation (in the limit of minimal accretion) was adapted from Hansen
(1999).  We have included a latent heat of crystallization of $0.77
kT$ per particle, but no fractionation.

 At early times in the evolution, the WD core temperature, $T_c$, is
much lower than the maximum temperature reached at the base of the
outer radiative envelope, $T_m$. In this limit, TB showed that $L$
becomes independent of $T_c$ and is only a function of $\timav$ and
$M$.  They found that the envelope state is determined self
consistently by balancing the heat released in the outer
(non-degenerate) envelope due to fluid motion down the entropy
gradient, $L_{\rm comp}\approx 2\timav kT_m/\mu m_p$, with that
carried through the envelope by radiative transfer $L\approx L_\odot
(T_m/10^8\ {\rm K})^{2.5}$, where $\mu=4/3$ is the mean molecular
weight in the He envelope, $k$ is Boltzmann's constant, and $m_p$ is
the baryon mass. This same state applies here when $T_c<T_m$ and we
have calculated this balance numerically using the quasi-static
envelope methods outlined in TB for He accretion and found,
\begin{equation}
L\approx 2.8\times 10^{-2}L_\odot \left(\timav\over 10^{-9} M_\odot {\rm
yr^{-1}}\right)^{1.4}\left(0.6 M_\odot\over M\right)^{0.3}. 
\end{equation} 
This relation holds for 
$T_c<  T_m\approx 2\times 10^7\ {\rm K}
(\timav/10^{-9}M_\odot {\rm yr^{-1}})^{0.5}$ and 
$10^{-10}M_\odot\ {\rm yr^{-1}} < \timav < 10^{-7}M_\odot\ {\rm yr^{-1}}$, 
although for our time histories $T_c\simeq T_m$ by the time
$\timav \simeq 10^{-9}M_\odot\ {\rm yr^{-1}}$ so that this form only
applies to Figure 1 for $\timav > 10^{-9}M_\odot\ {\rm yr^{-1}}$.
During this early
phase, the WD core is being heated as $\timav$ is
dropping.\footnote{The only other calculation we could
find that reported exiting luminosities are in Table 2 of Limongi \&
Tornambe (1991). While our value compares well to their calculations
with $M=0.6M_\odot$ as the starting WD mass, when $M=0.8M_\odot$ and
the accreted layer is thin, their values are as much as a factor of 50
lower for $\timav\sim 10^{-8}M_\odot$ yr$^{-1}$.  We believe that $L$,
which is largely due to compressional heat release in the outer layers
and was only of peripheral interest in their study, is not well
determined by their modeling for small accreted layer masses at
$M=0.85 M_\odot$.}

   As $\timav$ drops and $T_c$ increases, the accreting WD approaches the
equilibrium state where the outgoing luminosity is matched by the
internal energy release (Nomoto 1982; TB).  There is a brief
transition from core heating to cooling which appears in Figure
1 as a plateau in $T_c$.  In the equilibrium
state, a fairly good estimate of the compressional heating luminosity
is found by integrating throughout the $0.05M_\odot$ accreted 
degenerate He layer (identical to that
performed in Appendix B of TB, and Nomoto 1982), giving $L_{\rm comp}
\approx 10 \timav kT_c/\mu_{i,\rm He} m_p$, where $\mu_{i,\rm He}=4$
is the ion mean molecular weight in the helium envelope. As $\timav$
continues to drop, this will eventually become low enough that it is
exceeded by the core cooling luminosity $L_{\rm cool} = L_\odot
(T_c/10^8\ {\rm K})^{2.5}$.  At this point the WD evolution
\emph{decouples} from $\timav$ and cools much as an isolated object
would.  By setting $L_{\rm comp}= L_{\rm cool}$ and taking $T_c=
1$-2$\times 10^7$ K, we find $\dot M_{\rm dec}= 1$-$3\times
10^{-10}M_\odot$ yr$^{-1}$. These values are evident in Figure 1.

The precise value of $T_{c,\rm max}$ at the plateau depends on the
initial $T_c$ and when the He flashes cease, neither of which is well
known. This leads to some uncertainty in $L$ at the time of decoupling. However, analogous to isolated cooling WDs, the "initial" 
$T_{c,\rm max}$ is forgotten, and $L$ eventually only depends on the
time after decoupling, $\Delta t$.  This is simply the WD cooling age
from the time that heating ceased. By $10^9$ yr, $L\approx
10^{-3}L_\odot$, already consistent with a cooling DB WD of a similar
age (Hansen 1999).

Using the values tabulated by Limongi \& Tornambe (1991, hereafter LT91) for
accretion of Helium onto a $0.8M_\odot$ WD at a fixed rate, we find that our
determination of the heating time from the quasi-static model agrees well
with their  full numerical model.  Table 2 in LT91 gives a detail of $T_{\rm max}$
(the maximum with respect to radius) and $T_{\rm center}$ for a mass history
starting from $0.8M_\odot$ up to about $0.9M_\odot$.  We compare our $T_{\rm
core}=T_c(t)$ for a 0.9$M_\odot$ WD with a $0.05M_\odot$ Helium layer
accreting at a constant rate to $T_{\rm core} =(T_{\rm center}+T_{\rm
max})/2$ (representative of the heat content) from Table 2 of LT91, and find
that in both cases, for this range of mass, $dT_{\rm core}/dM$ is
approxmately constant.  We find $dT_{\rm core}/dM\approx 1.6\times 10^{8}$ K
$M_\odot^{-1}$ and $2.7\times 10^8$ K $M_\odot^{-1}$ for
$\timav=10^{-8}M_\odot\ {\rm yr}^{-1}$ and $5\times 10^{-8}M_\odot\ {\rm
yr}^{-1}$ respectively, compared to the values of $1.3\times 10^8$ K
$M_\odot^{-1}$ and  $2.7\times 10^{8}$ K $M_\odot^{-1}$ from LT91.  Also, the
$T_{c,\rm max}$ reached in the plateau ($\simeq 4\times 10^7$ K) reflects the
equilibrium $T_c$ for the $\timav$ at this epoch $\sim 5\times
10^{-9}M_\odot\ {\rm yr}^{-1}$.  This is consistent with the typical values
of interior temperatures, $\log T\simeq 7.5{-}7.7$, found by LT91 for a
$0.8M_\odot$ CO WD accreting Helium steadily at $\timav
\simeq5{-}10\times10^{-9}M_\odot\ {\rm yr}^{-1}$.

Before core crystallization begins, the WD's heat capacity is well
approximated by $C=3kM/\mu_i m_p$.  With the simple relation $L\approx
L_\odot (T_c/10^8\ {\rm K})^{2.5}(M/0.6M_\odot)$, $L=-CdT_c/dt$ gives
$L\approx 6\times 10^{-4}L_\odot t_9^{-5/3}(M/0.6M_\odot)$ where $t_9$
is time in Gyrs.  This agrees well between decoupling and 
the onset of crystallization, which occurs at $T_c=4(9)\times 10^6$ K
for $M=0.65(1.05)M_\odot$, corresponding to a time of
$1.4(0.4)\times 10^9$ yr and an orbital period of 60(45) minutes for
the hot donor evolution.  The effect of the latent heat release is
visible in the $T_c$ history and the heavier WD cools more quickly
at late times due to its earlier crystallization and the resulting
decrease in total heat capacity. The modifications due to latent heat
release change the $L-t$ relation to something closer to 
$L=6.8\times 10^{-4}L_\odot t_9^{-1.1}$ and $1.4\times 10^{-3}L_\odot t_9^{-1.5}$ for
$M=0.65M_\odot$ and $1.05M_\odot$ respectively in the 
1-8 Gyr region. By contrast, the accretion luminosity $L_{\rm acc}=GM\timav/R$ is a
smooth power law after about $10^7$ yr, given by $L_{\rm acc} =
8.3(19)\times 10^{-3}L_\odot t_9^{-1.3}$ for $M=0.65(1.05)M_\odot$.

\begin{figure}
\plotone{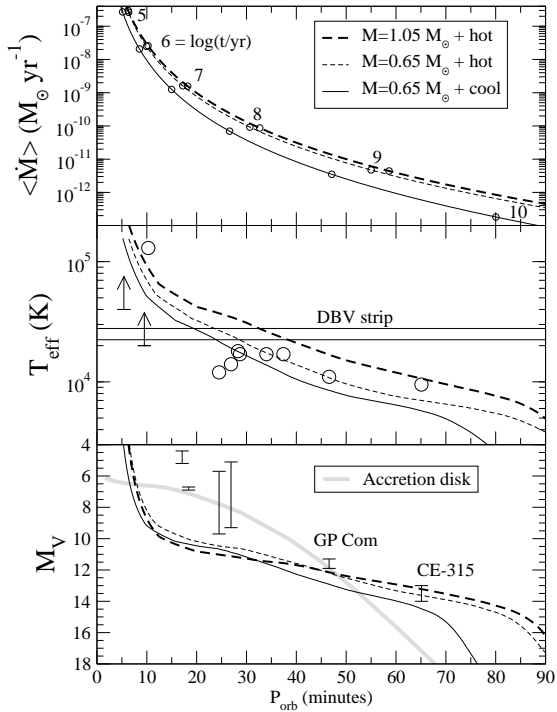}
\figcaption{
The time-averaged accretion rate, $\timav$, WD surface temperature,
$T_{\rm eff}$, and minimum absolute visual magnitude, $M_V$ for the
accreting WD in AM CVn binaries. The different WD curves are as in Figure
1 and the average accretion disk $M_V$ is shown by the
thick gray line (Nelemans et al. 2004).  The system age (time since
coming into contact) is indicated in the topmost panel by small circles
spaced evenly in $\log(t/{\rm yr})$ and demonstrates that a cool, more compact
donor leads to an older system at a given $P_{\rm orb}$.  In the $T_{\rm
eff}$ plot we delineate the DB instability strip (where isolated helium
rich WDs pulsate; Beauchamp  et al. 1999).  Constraints on $T_{\rm eff}$
for observed systems are shown by either lower limits or circles.  (See
Table 1 for a summary of measurements.) Some of the systems in the 20-40
minute range probably have a disk contribution even in quiescence,  giving a low $T_{\rm
eff}$ different from that of the WD surface.  GP Com and CE-315,
however, have both quiescent disks and measured distances, giving
$T_{\rm eff}$ and $M_V$ directly from the WD.  Comparing these to our
models we find that both of these systems appear to have hot donors.
}
\label{fig:compare} 
\end{figure}
  The bottom panel of Figure 1 shows the resulting $L(t)$, for which
there is very little contrast with the donor entropies.  However, when
displayed as a function of the observable, $P_{\rm orb}$ (see Figure 2), a contrast
immediately appears:  accretors in systems with hotter donors have
higher $T_{\rm eff}$'s.  The reason for this depends on the phase of
evolution, but is essentially related to the difference in speed of
evolution as set by the donor's $M$-$R$ relation.  In the
``compressional heating'' phase a lower entropy donor always produces
a lower $\timav$ at a given $P_{\rm orb}$, leading to a lower $L$.
By an orbital period of 30-40 minutes the WD is cooling and the time
since decoupling becomes larger than the uncertainty introduced by
the unknown $T_{c,\rm max}$, so that at longer orbital periods we can
safely predict  $L$.  Low-entropy donors reach the decoupling
$\timav$ at shorter orbital periods and evolve more slowly in $P_{\rm
orb}$, so that the WDs at a given $P_{\rm orb}$ always look "older"
(i.e.  colder) if their donor was initially of low entropy. The track
for the cool  donor (30th percentile) gives a good maximum $\Delta t$, which
for $P_{\rm orb}>40$ min can be fit by $\Delta t \approx 5\times
10^{8} {\rm yr} (P_{\rm orb}/40 {\rm min})^{4.37}$, and thus
yields the relation for the coldest possible $0.65M_\odot$ WD at these orbital
periods, $L_{min}\approx 1.5\times 10^{-3}L_\odot (40\  {\rm min}/P_{\rm orb})^{4.8}$. 

\section{Observability of the Accreting White Dwarf} 

 In order to assess the likelihood of observing the accreting WD, we
must compare its luminosity to that from active accretion. When the WD
is near equilibrium or simply emitting the compressional heating
luminosity, $L\sim kT_c \timav/ m_p$ will be much less than the
accretion luminosity, $L_{acc}\approx GM \timav/R$, since $kT_c\ll GMm_p/R$.
Seeing the hot WD then depends on either geometry (for example, an
edge-on systmem like the newly discovered eclipsing
binary SDSS~J0926+3624; Anderson et al. 2005) or 
probes of different spectral regions. Four classes among observed systems
are delineated in Table 1 by horizontal lines. The shortest period
systems are in a phase where the accretion stream directly hits the WD
(e.g.  Marsh \& Steeghs 2002; Israel et al. 2002), providing the first
opportunity.  In this case there is no disc and there are times when
the direct impact location is out of sight, providing an excellent
chance to see the (very hot) accreting WD.  At slightly 
longer $P_{\rm orb}$, around 20 minutes, all systems are expected to have a
consistently bright accretion disc in the optical/UV that would
energetically dominate the WD (e.g. Nelemans et al.  2004) as noted
above. Indeed, Nelemans et al. (2004) showed that a simple accretion
disc model at 15 minutes would give $M_V\approx 7$
and  temperatures of $\approx 30,000$ K, swamping the hot WD. This model is shown by the thick
grey line on the bottom panel of Figure 2. 
\begin{deluxetable*}{l c c c c c c  l}
\tablecaption{AM CVn System Parameters and Inferred Accreting WD
Properties\label{tab:systems}}
\tablehead{\colhead{System}&\colhead{$P_{\rm orb}$}&\colhead{$\timav_{\rm
min}$}&\colhead{$M_{c,\, {\rm min}}$}&\colhead{$T_{\rm eff}$}&\colhead{V
Range}&\colhead{$M_V$}&\colhead{Ref}\\
\colhead{Name}&\colhead{(min)}&\colhead{($M_\odot \
{\rm yr^{-1}}$)}&\colhead{($M_\odot$)}&\colhead{(1000 K)}&&&}
\startdata
RX~J0806&5.35&$1.7\times 10^{-7}$&0.125&$>40$&21.1&\nodata&1\\
V407 Vul&9.49&$ 7.0\times 10^{-9}$&0.068&$>20$&19.9&\nodata&2\\
\hline
ES Cet&10.3&$4.4\times 10^{-9}$&0.062&130$\pm$10&16.9&\nodata&3,17\\
AM CVn&17.1&$3.7\times 10^{-10}$&0.035&\nodata&14.0&4.8&4\\
HP Lib&18.4&$3.8\times 10^{-10}$&0.032&\nodata &13.6&7.0&4\\
\hline
CR Boo&24.5&$7.6\times 10^{-11}$&0.023&12&13.5-17.5&5.7--9.7&4,5,6,
7\\
KL Dra&25.0&$4.3\times 10^{-11}$&0.022&\nodata&16.8-20&\nodata&8\\
V803 Cen&26.9&$7.6\times 10^{-11}$&0.020&14&13.2-17.4&5.1--9.3&9\\
CP Eri&28.4&$3.5\times 10^{-11}$&0.019&17&16.5-19.7&\nodata&10,19\\
SDSSJ0926+3624&28.3&$3.5\times 10^{-11}$&0.019&18&18-19&\nodata&20\\
2003aw&33.8&$1.4\times 10^{-11}$&0.015&17&16.5-20.3&\nodata&11,18\\
\hline
SDSSJ1240-01&37.4&$5.6\times 10^{-12}$&0.013&17&19.6&\nodata&12\\
GP Com&46.5&$2.2\times 10^{-12}$&0.010&11&15.7-16&11.3--11.7&4,13,14\\
CE-315&65.1&$3.4 \times 10^{-13}$&0.006&9.5&17.6&13.2&15,16 
\enddata
\tablerefs{
1 Israel et al., 2002;
2 Ramsay et al., 2002;
3 Warner \& Woudt, 2002;
4 Groot et al., 2005, in prep.;
5 Wood et al., 1987;
6 Provencal et al., 1997;
7 Patterson et al., 1997;
8 Wood et al., 2002;
9 Patterson et al., 2000;
10 Abbott et al., 1992;
11 Warner \& Woudt, 2003;
12 Roelofs et al., 2005a;
13 Thorstensen, 2003;
14 Warner, 1995;
15 Thorstensen, Priv. Comm.;
16 Ruiz et al. 2001;
17 Espaillat et al. 2005;
18 Roelofs et al, 2005b;
19 Sion, Priv. Comm.;
20 Anderson et al. 2005}
\end{deluxetable*}

  The third group of AM CVn systems, with again slightly longer
orbital periods, show large brightness variations due to a thermal
instability in the disc (Tsugawa \& Osaki 1997). Although at this
$\timav$ the disc absolute magnitude is expected to be brighter than
the WD, the low state is a time when the accreting WD can be detected, 
just as in the hydrogen accreting CVs (Sion 1991; Townsley \& Bildsten 2003). 

  The last group, with the longest orbital periods, again show hardly
any brightness variations. In the context of the thermal instability
interpretation, these systems could be on the lower, stable, branch of
the instability curve and thus would be cool discs, consisting mainly
of neutral helium. The steadily accreting luminosity is only $\sim 10$
times larger than the cooling WD (see \S 2), enhancing the prospects
of seeing the accreting WD relative to shorter orbital periods, where
the contrast is greater. In addition, the disc temperature will
generally be cooler than the WD as shown in the bottom panel of Figure 2.

\subsection{Non-Radial Oscillations of the Accreting WD} 

In addition to directly seeing the re-heated WD in the AM CVn
binaries, there are also opportunities for seismological tests as the
WD passes through the non-radial pulsation instability strip ($T_{\rm
eff}=22,400-27,800\ {\rm K}$ for isolated DBs; Beauchamp et al. 1999),
which we plot in Figure 2. The pulsation periods are
200-1000 seconds (see Handler 2003 for a recent summary) and are
typically $l=1,2$ g-modes of high radial order (e.g. Kotak et al. 2003).

We do not know, {\it a priori}, that the instability strip will be the
same for a He accreting WD. The clear structural difference in the
accreting WD is the mass of the helium envelope, which will likely
exceed $0.05M_\odot$ and be seismically measurable. It seems unlikely
that the thicker He envelope would alter the instability strip, since
the relevant "pump" for the mode excitation occurs at low pressures
(see Gautschy \& Althaus 2002 for a recent study). However, active
accretion onto the outer atmosphere might make a difference,
especially if the convection zone (responsible for the mode
excitation; e.g. Brickhill 1991) is altered. It also seems  possible that the mode 
pump could become active during the cooling of the outer WD layers (Piro, Arras \& Bildsten 2005) 
following an accretion disk outburst. Indeed, 
Wood et al. (1987) reported excess power at periods of
100-200 seconds when CR Boo had an optical magnitude of $V=15.4$,
which might well be the passing of the accreting WD through the instability
strip following an accretion outburst.

 The most dramatic alteration of the g-mode frequencies would be from
WD rotation at periods comparable to the g-mode periods, as the WD has
accumulated a significant amount of mass and angular momentum since
the last mass ejecting He novae event.  This angular momentum either
resides solely in the He layer or has been shared with the underlying
C/O WD in the $\sim 10^8$ years spent at the $P_{\rm orb}\approx
20-40$ minutes where $T_{\rm eff}$ is in the isolated DB instability
strip (see Figure 2).  Motivated by the likelihood
that very weak internal magnetic fields can easily mediate angular
momentum transport on this timescale (e.g. Spruit 2002), we presume
that all accreted angular momentum after the last He novae event is
shared with the whole WD, yielding a lower limit to the rotation
frequency of the He shell, where much of the g-mode energy resides.

We assume that helium accretes onto the WD with the specific angular
momentum of the inner edge of the accretion disk, $(GMR)^{1/2}$, and
that the amount of accreted matter, $\Delta M\ll M$. The final WD spin
frequency is then $2\pi/P_{\rm spin}=(GMR)^{1/2}\Delta M/I$, where
$I\approx 0.2MR^2$ is the WD moment of inertia, yielding $2\pi/P_{\rm
spin}= 5\Delta M(G/MR^3)^{1/2}$,
\begin{equation}
\label{eq:pspin} 
P_{\rm spin}\approx 50\  {\rm s}\left(0.05 M_\odot\over \Delta
M\right)\left(M\over 0.6M_\odot\right)^{1/2}\left(R\over 9\times 10^8
{\rm cm}\right)^{3/2},
\end{equation}
which reaches about 100 seconds when $M=1.2M_\odot$. Hence, it appears
very likely that the WD surface layer (and maybe even core) is
rotating rapidly enough (i.e. $P_{\rm spin}<P_{\rm mode}$) that the
seismology will be dramatically altered from the isolated DB case.
The discovery and study of a DB pulsator in an AM CVn binary would
provide an unprecendented opportunity to measure the accumulated He
layer mass on the C/O WD and the angular momentum transport mechanisms
within it. The most probable orbital periods for this range from 20 to
35 minutes.

The rapid surface rotation will broaden any spectral features originating there, 
as the resulting minimum equatorial rotational velocity is
$\approx 1250 \ {\rm km \ s^{-1}}(\Delta M/0.05 M_\odot)$, nearly
independent of $M$. The narrow, central line emission (i.e. spikes) seen in  
He lines of  GP Com, CE 315 and SDSS1240 (Marsh 1999, Ruiz 
et al, 2001, Roelofs et al, 2005a) are generally believed to originate on the 
accreting WDs. However, their widths are much less than this 
estimate, suggesting either a different origin or an effective loss of angular 
momentum from the rapidly spinning accretor.

\subsection{ Atmospheric Properties} 

Another difference with single DB WDs is that $^{14}$N is present in
the accreting material at a mass fraction of 1-2\% (due to CNO burning
in the progenitor star of the He donor). This $^{14}$N would rapidly
sink out of the WD photosphere were it not for the constant mixing
provided by the underlying convection zone of mass $M_{\rm conv}$. It
is thus the sedimentation time, $t_{\rm sed}$, at the base of the
convection zone that sets the relevant timescale. Paquette et al.
(1986) calculated $t_{\rm sed}$ for $^{14}$N in pure He atmospheres of
isolated WDs, and defined an $\dot M_{\rm sed}=M_{\rm conv}/t_{\rm
sed}$, below which the $^{14}$N in the convective envelope will be
less abundant than in the accreting material. For $M=0.6M_\odot$ WDs,
$\dot M_{\rm sed}\approx 10^{-12}M_\odot {\rm yr^{-1}}$. Though
$M_{\rm conv}$ decreases rapidly with increasing $M$ (see Fontaine \&
Van Horn 1976), so does the sedimentation time, making the actual
value of $\dot M_{\rm sed}$ not too dependent on $M$ for the range of
$T_{\rm eff}$ relevant here (see Figure 14 in Paquette et
al. 1986). More recent calculations with updated opacities (MacDonald,
Hernanz, \& Jose 1998) found larger $M_{\rm conv}$ values by factors
of 4-6 from Paquette et al. (1986). Hence, for $P_{\rm orb}<40 $
minutes, the atmospheric N abundance will be comparable to that in the
accreted material. At longer orbital periods, depending on the state
of the donor, it seems plausible that $\timav$ will be less than $\dot
M_{\rm sed}$, and the $^{14}$N will tend to sediment. Detection and
study of atmospheric N (i.e. not from the accretion disk) in GP Com,
CE-315 and SDSS J1240 could thus constrain the sedimentation physics.

 The abundant presence of $^{14}$N may simplify the spectral modeling
of these WDs relative to what has been done for pure He atmospheres
(see Bergeron, Saumon \& Wesemael 1995a). This is because the low
first ionization potential of $^{14}$N will provide free electrons
more readily than the He. This may well allow the ${\rm He}^-$
free-free opacity to more readily dominate for the relevant
$T_{\rm eff}$'s,  but only calculations will tell. This could also alter
the location of the pulsational instability strip.

\section{Comparison to Observations}

We tabulate relevant observed and inferred properties of the AM CVn
binaries in Table 1.  The minimum donor mass, $M_{c, \rm min}$, for
each system is the mass of the Deloye et al. (2005) fully degenerate
donor model that fills the Roche lobe at the system's $P_{\rm orb}$.
The minimum $\timav_{\rm min}$ at fixed $P_{\rm orb}$ occurs when $M_c
= M_{c,\,{\rm min}}$ (Deloye et al 2005).  We thus take $M_{c,\,{\rm
min}}$ and calculate the tabulated values of $\timav_{\rm min}$ as
follows.  For systems in which the mass ratio has been inferred (all
the AM CVn binaries \emph{except} RX~J0806, V407 Vul, ES Cet, and
SDSS~J0926+3624), we use it and $M_{c,{\rm min}}$ to find $M$ and thus
$\timav_{\rm min}$.  When there is no mass ratio constraint, we assume
that $\timav_{\rm min}$ is given by the minimum possible value, which
occurs at a mass ratio of $\approx 0.3$ (this determined the entries
for RX~J0806, V407 Vul, and ES Cet), except for SDSS~J0926+3624, for
which we use the mass ratio of CP Eri.

Our discussion in the previous sections outlined the 
 opportunities for detecting the hot WD and we now discuss in more
detail the implications of our calculations in regards to
\emph{directly} observing the WD in these binaries.

\subsection{Short-period systems}

The evolutionary state of the two shortest period systems,
RX~J0806.3+1527 ($P_{\rm orb}=5.36$ min) and RX~J1914.4+2456
($\equiv$V407 Vul, $P_{\rm orb}=9.48$ min) is uncertain and different
models have been proposed (see e.g. Wu et
al. 2002; Ramsay et al. 2002; Marsh \& Steeghs, 2002; Marsh \&
Nelemans, 2005; Dall'Osso et al. 2005; Willems \& Kalogera 2005). 
Marsh \& Steeghs (2002) suggested that V407 Vul is a
direct-impact accretor where the accretion stream directly strikes the
accreting WD. Noting the earlier work of Townsley \& Bildsten (2002),
they suggested that much of the optical light could originate from 
the hot, accreting WD. 

Our high $\timav$ calculations
allow us to address the direct impact scenario. Even though the accreting material arrives
over a limited part of the WD, we expect complete spreading by the
time it has reached depths relevant for compressional heating (Piro \& Bildsten 2004), thus
allowing our calculation to apply. At the minimum $\timav\approx
10^{-8}M_\odot \ {\rm yr^{-1}}$ expected for V407 Vul (see Table 1), we
predict that $L\approx L_\odot$.  The predicted range of 
$T_{\rm eff}\approx 50,000-100,000 \ {\rm K}$ 
and absolute magnitude, $M_V\approx 8$. 
The analysis of V407 Vul is complicated by the
fact that the optical light is dominated by a G9V star (Steeghs et
al., in prep). However, the variable part of the light clearly shows a
blue color, with implied effective temperature $>20,000$ K (see Table
1), consistent with our calculations. The estimated dereddened visual
magnitude is 16.9 (Ramsay et al., 2002) which includes a substantial
G-star contribution. Combined with the estimated $M_V\approx 8$, 
this suggests a distance greater than 500 pc if V407 Vul is a direct impact 
accretor. 

For RX~J0806.3+1527 ($P_{\rm orb}=5.36$ min), the situation is quite different.
The shorter orbital period implies a luminosity of $\approx 44L_\odot$,
$T_{\rm eff}\approx 140,000$K and $M_V\approx 4.7$ for the accretor. The
estimated $T_{\rm eff}$ agrees  with the observed very blue spectrum, for which
Israel et al. (2002) find a lower limit of 40,000 K. The apparent magnitude of
21.1 would put the system at 20 kpc, 8 kpc above the Galactic plane.
A similar distance is needed to interpret the rather low X-ray
flux (Israel et al. 2002) as being generated by $\timav$. 

It was recently suggested that the low X-ray flux could be reconciled with much
smaller distances  due to a temporarily lower $\dot M\ll \timav$ or due to most
of the accretion luminosity being emitted in the UV (Marsh \& Nelemans 2005;
Willems \& Kalogera, 2005). However, our calculations show that not only the
X-ray luminosity, but also the optical flux tracks $\timav$, so that the latter
suggestion would still produce a much too bright WD.  The compressional heating
luminosity would track a drop in $\dot M$ on the thermal time of the accreted
layer, $\tau \sim c_P T_m (0.05M_\odot)/L \sim 10^6 {\rm yr} (\timav/
10^{-7}M_\odot\ {\rm yr^{-1}})^{-0.9}$ where $T_m$ is the maximum temperature,
and $L$ is given by equation (1). This timescale is much longer than the
timescale for which a temporarily lower $\dot M$ can be sustained ($\sim100$
yrs, Marsh \& Nelemans 2005), leading to a bright WD even at low $\dot M$.
Therefore, for RX~J0806.3+1527 to be a direct impact accretor, the only
solution seems to be a very large distance.

The third system in the table, ES Cet, is clearly a helium accreting
WD. Espaillat et al. (2005) suggest it is a direct impact
accretor and find $T_{\rm eff}\approx 130,000$ K and radius
 $R\approx 4.5 \times 10^8$ cm for a distance of 300 pc. They interpret this
as the size of the impact spot, but our calculations suggest that the whole WD should
be very hot. This observed temperature is in the expected range, and
the radius either suggests a massive ($1.1 M_\odot$) WD 
or  a less massive WD at a larger distance than they assumed. 

\subsection{Outbursting sources}

 For the outbursting systems the spectra are dominated by the accretion disc
when they are in their high state. The absolute magnitudes of the disc are at
least 6 -- 8, much brighter than the expected absolute magnitudes of the
accreting WDs of about 11 (Figure 2). However, in the low state,
the WD can contribute significantly to the spectra, potentially producing the
observed continuum. Between 20 and 30 min, the expected disc absolute magnitude
drops, while the WD $M_V$  stays roughly constant. Interestingly,
the magnitude range observed in these outbursting systems decreases with
increasing orbital period. While CR Boo has a range of 5 magnitudes, V803
Cen is less, at 4.2, and CP  Eri only changes by 3.2 magnitudes.

To check if the low-state spectra could be dominated by the  WD, we estimate the temperature from the shape  of the 4000--6000 \AA\  continuum
from the few spectra that are available.  We assume a black-body
spectrum, which reasonably produces the broad band colors of
theoretical DB white dwarf spectra (Bergeron, Wesemael \& Beauchamp 1995b). 
Our temperature determinations in Table 1 are
accurate at about the 10\% level\footnote{For $T_{\rm eff} \la 17000$ K the
black body colors overestimate $T_{\rm eff}$ by 500--1000 K, while above that
temperature they underestimate the temperature by a similar amount. In
addition, in the region considered here, the change in the slope of the black
body spectrum decreases with temperature above $\sim$15000 K, adding to the
inaccuracy at higher temperatures}. For CR Boo and V803 Cen, the absolute
magnitudes are also known (Groot et al. in prep, see Table 1). 

For CR Boo and V803 Cen, the temperatures are below the expected WD
temperatures and indeed the absolute magnitudes in the low state are about 1
magnitude above the predicted WD brightness (see Figure 2), suggesting a
significant contribution from the disc (which is cooler than the WD, but of larger 
area), even in the low-state. Using the slope of spectrum in the low
state, we found a temperature of $30,000 {\rm K}$ for CP Eri, 
whereas a recent UV spectrum suggests $17,000 {\rm K}$ (Sion, priv.
comm.). 
We have to be careful with these low-state estimates, though, as a second
low-state spectrum of CR Boo (Patterson et al. 1997) shows a very different
continuum shape, suggesting a cool or hybrid cool plus hot spectrum.

The recently discovered eclipsing AM CVn binary, SDSS~J0926+3624, at
$P_{\rm orb}=28.3 \ {\rm min}$ (Anderson et al. 2005) provides an
opportunity for probing the distinct contributions of the hot WD and
the disk. Anderson et al. (2005) report that the $\approx 1$ minute
duration of the eclipse is consistent with the $0.02M_\odot$ secondary
transiting a hot spot or the inner disk. It is also consistent with an
eclipse of the hot WD. Fast photometry and phase resolved spectroscopy could
confirm that the WD contributes and would tie the eclipse minimum to the WD
rather than the disk, yielding a more robust measurement of the orbital
period change expected from gravitational wave driven mass transfer (Anderson
et al.  2005).

\subsection{Long-period systems}

 For GP Com and CE 315 we derived the temperatures in the same way as
described above and interpret the temperature as that of the accreting
WD. For GP Com, using the known distance (d = 67 pc; Thorstensen
2003), we can derive the emitting area. Thorstensen (2003) gives a
typical observed V magnitude of 16.1, but Warner (1995) lists the
range 15.7--16.0. We therefore use V=15.7--16.1 as the observed range.
Together with $m - M = 4.2\pm0.2$ (Thorstensen 2003) this gives $M_V =
11.7\pm0.4$. From Bergeron et al. (1995b) we use the bolometric
correction for a 11,000 K DB model, being -0.57. This gives a
bolometric magnitude $M_{\rm bol} = 11.13 \pm0.4$, i.e. a luminosity
of $2.78 \times 10^{-3} L_\odot$ with an error of $44 \%$. Together
with the effective temperature, this gives the size of the emitting
region of $10^9$ cm, i.e. a typical WD radius!

 Using the same argument for CE 315, with a distance of 77 pc
(Thorstensen, priv. comm.), using a effective temperature of 9500 K and a
bolometric correction of -0.34, we find an emitting region of $6
\times 10^8$ cm, implying a more massive WD of $\approx 0.9
M_\odot$. Indeed, its absolute magnitude falls close to the track for
a 1.0 $M_\odot$ accretor (Fig. 2).  Interestingly, two
systems have been recently discovered that, in their low state,
clearly show the broad absorption lines of a DB white dwarf, with
estimated temperature of 17,000 K (SDSS J1240-01 and SN2003aw, Roelofs
et al. 2005a; Warner \& Woudt, 2003, Roelofs et al,
2005b). Based on an absolute magnitude of 11 (Fig. 2),
these systems should have distances of 735 pc (SN2003aw with V = 20.3)
and 525 pc (SDSS J1240-01, with V = 19.6).

\section{Summary and Future Work}

 We have shown (both from theory and observation) that thermal
emission from the accreting WD within an AM CVn binary is an important
contributor to the observed optical emission. For the direct impact
systems with the shortest orbital periods, the hot WD is visible when
the hot spot is out of the line of sight. At longer orbital periods,
the WD can either be seen when the disk is in a low state, or at even
longer orbital periods ($P_{\rm orb}>40 $ min), the WD always dominates the
accretion luminosity in the optical bandpass. 

We compared our theoretical results to the temperature and brightness
of observed systems.  The accreting WD is the dominant contributor to
the optical light for the direct impact accretors and constrains their
distances.  In particular, the expected brightness of the WD is so
large for RX~J0806 that it can only be in the direct impact phase if
it is located at a distance of 20 kpc, far outside the Galactic
disc. For $P_{\rm orb}=20-40 {\rm min}$, our predicted absolute
magnitude of the WD is $M_V\approx 11$. This allows for estimates of
the distance to the two newly discovered AM CVn systems (2003aw and
SDSSJ1240-01) in which the DB white dwarf is visible.  The two
outbursting systems (CR Boo and V803 Cen) with parallaxes have
$M_V\approx 10$ in quiescence, 1 magnitude brighter than our estimate
for the WD. This suggests that even in the low state there is a disc
contribution. For GP Com and CE-315, a combination of estimated
effective temperature and known distance suggests that the optical
continuum emission is dominated by emission from the WD, in agreement
with our calculations.  For these two systems we find consistency with
a hot (high-entropy) donor.  Whether this donor was born hot and has
evolved at fairly constant entropy or is heated by ongoing processes
is unknown but could be addressed by future theoretical work.

Our successful comparisons to observations gives us confidence to
apply our work to predict just how faint these binaries can become.
For example, either mass WD reaches $M_V=15$ at 5 Gyrs, and at 10
Gyrs, the absolute magnitudes are $M_V>20$. This will allow for
meaningful constraints on the population density of such binaries from
deep optical searches, both in the field (e.g. the 3 additional
candidates found by SDSS, Anderson et. al. 2005) and in Globular
Clusters.

 Much work remains to be done on the theoretical side, the most
 important of which is to resolve better the nature and occurrence
 rate of He flashes at early stages in the evolution of these
 binaries. In addition to better constraining the 
 remaining He mass on the accreting WD at late times, it would also
 allow us to predict the He novae rate and potentially constrain the
 population from the measurements of such events. Perhaps the most
 dramatic development would be the secure detection of non-radial
 oscillations in these WDs, allowing for measurements of 
both the WD rotation rate and accumulated He layer mass.

\acknowledgments

 We thank Tony Piro for comments on the manuscript, and Ed Sion for
helpful input as a referee.  This work was supported by the National
Science Foundation (NSF) under grants PHY99-07949, AST02-05956 and
AST02-00876.  D.M.T. is supported by the NSF Physics Frontier Centers'
Joint Institute for Nuclear Astrophysics under grant PHY02-16783 and
DOE under grant DE-FG 02-91ER 40606. GN is supported by NWO Veni grant
639.041.405.



\end{document}